\documentclass[12pt]{article}

\usepackage{amsfonts}

\usepackage{amsmath}
\usepackage{amssymb}
\usepackage{float}
\usepackage{titlesec}

\titleformat*{\section}{\large\bfseries}
\titleformat*{\subsection}{\bfseries}

\newtheorem{prop}{Proposition}[section]

\newtheorem{theorem}{Theorem}[section]
\newtheorem{remark}[theorem]{Remark}

\newtheorem{lemma}[theorem]{Lemma}

 \def\2{I$\!$I}

\newcommand{\1}{1\!\!1}

\def\*{{\phantom *}}

\newcommand{\R}{\Bbb R}

\newcommand{\bml}[1]{\begin{multline}\label{#1}}
\newcommand{\bee}{\begin{equation}}
\newcommand{\bed}{\begin{displaymath}}
\newcommand{\ee}{\end{equation}}

\newcommand{\bs}{\begin{split}}

\newcommand{\be}{\beta}
\newcommand{\ga}{\gamma} \newcommand{\Ga}{\Gamma}
 
\newcommand{\la}{\lambda} \newcommand{\La}{\Lambda}

\newcommand{\si}{\sigma}

 \newcommand{\sm}{\setminus}

\newcommand{\w}{\widetilde}

\begin{document}
\large

\begin{center}
\renewcommand{\thefootnote}{\fnsymbol{footnote}}
{\Large \textbf{Virial Expansions for Correlation Functions in Canonical Ensemble}}

\vspace{0.5cm}
{\textbf{ A.~L.~Rebenko}}
\date{}
\begin{footnotesize}
\begin{tabbing}
 Institute of Mathematics, Ukrainian National Academy of Sciences, Kyiv, Ukraine.
\end{tabbing}
\end{footnotesize}


\setcounter{footnote}{0} \renewcommand{\thefootnote}{\arabic{footnote}}


\vspace{0.5cm}
\textbf{Abstract} \\[0pt]
\end{center}
The infinite set of coupled integral nonlinear equations for correlation functions in the case
of classical canonical ensemble    is considered.
Some kind of graph expansions of correlation functions  in the density parameter are constructed.
Existing of unique solutions for small value of density and high temperature is discussed.
\vspace{0.5cm}


\medskip

\textheight 24.0cm \textwidth 16.0cm \headheight 0cm \headsep 0cm \footskip %
1cm \topmargin 0cm



\section{Introduction.}

\setcounter{equation}{0} \renewcommand{\theequation}{\arabic{section}.%
\arabic{equation}} 

The first mathematically rigorous studies of infinite statistical systems were the works of Bogolyubov \cite {Bog46}
in the canonical ensemble and Ruelle \cite{Ru63, Ru69} in the grand canonical ensemble.
In the monograph \cite {Bog46} Bogolyubov developed a theory
of expansions in degrees of density, which, although not yet rigorously mathematically justified,
but was based on the fundamental principles of statistical mechanics and included all
known at that time expressions of Ursel-Mayer theory on the basis of his derived integro-differential
equations for $ m $ -particle distribution functions. Really rigorous result was formulated in \cite{BogKh49}
(see details in \cite{Kh56-1, Kh56-2}), but his description covered only non-negative
interaction potentials of the particles.
It was only in 1969 that Bogolyubov, Petrina, and Khatset \cite{BPH69} succeeded in generalizing these results
to the case of stable regular potentials.
Following Ruelle's operator method, they derived a recurrent relation, which, in the thermodynamic limit,
 turned into the Kirkwood-Salzburg equation.
But the role of activity $z$  is played  some function of the density $ a(\varrho) $, which is determined
by series in the set of all distribution functions.
This essentially means that the equation is an infinite system of nonlinear equations,
and its iterations are not a series expansion in the density.
A little later, these equations were derived by Pogorelov \cite {Po75} directly
from Bogolyubov's integro-differential
equations (see \cite {Bog46}) and a theorem  on the existence of a unique solution
(as of nonlinear operator equation) for small values of density  was proved.

The virial expansion is most often understood as the expansion of pressure in terms of the density \cite{MaMa40}
and obtained, by formally inverting the density as a power series of the fugacity and
composing the latter with the expansion for the pressure (Mayer's   power series in the fugacity).
In addition, we work with expansions of correlation functions in the thermodynamic limit.
Existence of a thermodynamic limit for  correlation functions in a canonical ensemble
"is more subtle than for thermodynamic quantities like pressure and free energy which are on a logarithmic
scale" \cite{KuTs18}. One more thing, the terms of the Mayer's  cluster expansions can be expressed in terms of forest-graphs
(see, i.e. \cite{ Ru69, MP77, Re2021}, while those of the virial expansion are usually
written in terms of 2-connected diagrams (see, for instance, the classical reference \cite{UF63}).

In this article we construct first an expansion for correlation functions by some special type of forest-graphs,
 but with huge number of terms in every order of density. Due to the fact that this series consists of two parts
 that have opposite signs, it is possible to reduce the contributions and, as a result, obtain its representation by
 some set of graphs, connected components of which are 2-connected graphs. Such set of graphs was also defined
in the article of Jansen \cite{Ja21} (see sect. 2).

The well-known proofs of convergence for the virial expansion are based on the above mentioned
inversion. Obviously the first result goes back to the work of Lebowitz and Penrose \cite{LePe64}
and some alternative approaches were given by Groeneveld \cite{Gro67}  (see also \cite{PT12, RT15, NF20},
which was continued in a more general and modern presentation in the work \cite{Ja21}.
The present work apparently has quite a few points of contact with the work \cite{Ja21}.

The article is organized as follows. In Section 2 we give a short mathematical introduction of
concepts and formulas in which the presentation of the article is carried out.
In Section 3, we briefly show the derivation of equations for correlation functions in the canonical ensemble.
The fourth section is devoted to the representation of the solution in the form of the Lebesgue-Poisson integral
of some kernel, for which we write the recurrent nonlinear equation. And finally, in the last section,
the question of its solution in the form of an expansion in terms of powers of density and
the convergence of the integral representation for correlation functions are discussed.

\section{On some mathematical concepts}

\subsection{Configuration spaces and Lebesgue-Poisson measure.}
Let $ \sigma $ be a Lebesgue measure in $ \R^d $. {\it Configuration space} $ \Ga \;: = \; \Ga_{\R^d} $ 
consists of all locally finite subsets of the space $\R^d $.
This definition is quite natural from the point of view of physics, because
an infinite number of particles cannot be in a bounded volume.

Denote the set of all finite configurations of the space  $\Ga $ by $ \Ga_0 $, and by $ \Ga^{(n)} $
configuration space with a fixed number of points.

If all such configurations are in some bounded set $ \La\subset\R^d $,
then the corresponding space will be $ \Ga^{(n)}_\La $.

State of an ideal gas in equilibrium statistical mechanics
is described by the {\it Poisson} measure  $ \pi_{z \sigma} $ on
configuration space $ \Ga $, where $ z> 0 $ is the activity (physical parameter that
associated with the density of particles in the system).

But we need  so-called Lebesgue-Poisson measure, i.e. the Poisson measure
on the space of finite configurations, which in some sense is absolutely continuous
with respect  to the Lebesgue measure in contrast to the $ \pi_{\sigma} $ measure for $ \Ga $. Integral for
Lebesgue-Poisson measure in space
$ \Ga_0 $ (or $ \Ga_\Lambda $) is determined by the formula:
\begin{align}\label{IL-P}
\int_{\Ga_X}F(\ga)\la_{z\si}(d\ga)\;&:=\;\sum_{n=0}^\infty\frac{z^n}{n!}\int_X\cdots\int_X
F(\{x_1,...,x_n\})\si(dx_1)\cdots\si(dx_n)=\notag\\
&=\sum_{n=0}^\infty\frac{z^n}{n!}\int_X\cdots\int_X
F_n(x_1,...,x_n)dx_1\cdots dx_n,
\end{align}
for all measurable functions $ F = \{F_n \}_{n \geq 0} $, $ F_n \in L^{1}(X^n) $, and
  $ \Gamma_X \in \{\Gamma_0, \Gamma_\Lambda $).
  Such  measure not only simplifies the recording of cumbersome expressions and combinatorial proofs,
  but also allows  to get some new results based on properties
  infinite divisibility (see, for example, \cite{Re93, ReSh97, Re98, PR07}.

In the language of integrals with the measure $\la_{z\sigma}$ we give one well-known and important identity,
which will be used very often below (see, for example, \cite{Ku99}).

\begin{lemma}\label{lem2}
 For all measurable functions $G:\Ga_0\mapsto \R$ and $H: \Ga_0\times\Ga_0\mapsto \R$,  for which
 $G(\xi\cup\ga)H(\xi,\ga)\in L^1(\Ga_0\times\Ga_0, \la_{\si}\otimes\la_\si) $, the following equality is true:
\begin{equation}\label{EgLem2}
\int_{\Ga_0}G(\ga)\sum_{\xi\subseteq\ga}H(\xi,\ga\sm\xi)\la_\si(d\ga)=\int_{\Ga_0}\int_{\Ga_0}G(\xi\cup\ga)H(\xi,\ga)\la_{\si}(d\ga)\la_\si(d\xi).
 \end{equation}
\end{lemma}
{\it Proof}.
 Let
 $\xi\restriction\Ga_0^{(k)}=\{x_1,...,x_k\}:=\{x\}_1^k$,
 $\ga\restriction\Ga_0^{(m)}=\{x_{k+1},...,x_{k+m}\}:=\{x\}_{k+1}^{k+m}$. Then
\begin{align*}
&\int_{\Ga_0}\int_{\Ga_0}G(\xi\cup\ga)H(\xi,\ga)\la_{\si}(d\ga)\la_\si(d\xi)=\\
&=\sum_{k,m=0}^\infty\frac{1}{k!m!}\int_{\R^{dk}}\int_{\R^{dm}}
G(\{x\}_1^{k+m})H(\{x\}_1^k,\{x\}_{k+1}^{k+m})\si(dx)^{k+m}=\\
&=\sum_{n=0}^\infty\frac{1}{n!}\sum_{k=0}^n {n \choose k}\int_{\R^{dn}}
G(\{x\}_1^{n})H(\{x\}_1^k,\{x\}_{k+1}^{n})\si(dx)^{n}=\\
&=\int_{\Ga_0}G(\ga)\sum_{\xi\subseteq\ga}H(\xi,\ga\sm\eta)\la_\si(d\ga).
\end{align*}
\hfill $\blacksquare$

Note that the introduction of the measure \eqref{IL-P} and the application of the lemma \ref{lem2} greatly simplifies the calculations of many combinatorial statements.

\subsection{Algebra of generating functionals.}

For any function
$\psi\in C_0(\Gamma_0)$ define  {\it generating functional }:
\begin{align}\label{GFf}
\w{F}_\psi(j)&\;=\;\sum_{N=0}^{\infty}\frac{1}{N!}\int_{\R^{dN}}(dx)^{N}j(x_1)\cdots j(x_N)\psi(x_1,\ldots, x_N)=\notag \\
&\;=\;\int_{\Gamma_0}e(j;\gamma)\psi(\gamma)\lambda_\sigma(d\gamma),
\end{align}
where functional $e(j;\gamma)$ is defined by the formula
\begin{align}\label{I-lam4}
e(j;\gamma):=\;&\begin{cases}
 1,\;&\gamma=\emptyset,\\
  \prod_{x\in\gamma}j(x),
\;&\gamma\in\Ga_0\sm\{\emptyset\},
  \end{cases}
\end{align}

 and $j\in C_0(\R^d)$ is bounded (take for simplicity  $j\leq 1$), continuum, nonnegative  function.

It is easy to calculate using the lemma \ref{lem2} that
\begin{equation}\label{odprod}
  \w{F}_{\psi_1}(j)\w{F}_{\psi_2}(j)\;=\;\w{F}_{\psi_1*\psi_2}(j)=\int_{\Gamma_0}e(j;\gamma)(\psi_1*\psi_2)(\gamma)\lambda_\sigma(d\gamma),
\end{equation}
where $\psi_1*\psi_2$ is product $\psi_1$ and $\psi_2$ in commutative algebra $\mathcal{A}$,
which was introduced by Ruelle (see \cite{Ru69}, Ch. 4). Formula \eqref{odprod}
establishes an unambiguous correspondence between the algebra of generating functionals with operation of ordinary multiplication
and the algebra $ \mathcal {A} $. Recall that for
$\gamma=\{x_1, x_2, ... x_n\}$
\begin{equation}\label{odprodRu}
(\psi_1*\psi_2)(\gamma)= \sum_{\xi\subseteq\gamma}\psi_{1}(\xi)\psi_{2}(\gamma\setminus\xi).
\end{equation}

\section{Description of the system of interacting particles in the canonical ensemble.}

\subsection{Interaction between particles}
We  consider the 2-particle interaction, which is described by the potential
$V_2(x,y)=\phi(|x-y|),\;\phi(0)=+\infty$. For any configuration $\ga\in\Ga_0$
an interaction energy  is
\begin{equation}\label{U}
U(\gamma)=U_{\phi}(\gamma):=\sum_{\{x,y\}\subset\gamma}\phi(|x-y|),
\end{equation}
and interaction $W(\eta; \gamma)$ between particles of configuration $\eta\in\Ga_0$ with particles of
configuration $\ga\in\Ga_0$ is
\begin{equation}\label{W}
W(\eta; \gamma):=\sum_{\substack{x\in\eta \\
y\in\gamma}}\phi(|x-y|).
\end{equation}

We impose the following conditions on the potential of interaction:

{\bf (A)}: 1. {\it Stability:}
\begin{equation}\label{stU}
U(\gamma)\;\geq\;-B|\gamma|,\;B\,\geq 0,\;\gamma\in\Gamma_0,
\end{equation}
\qquad  2. {\it  Regularity:}
\begin{equation}\label{reg}
C(\beta)\;=\;\int_{\R^d}dx|e^{-\beta\phi(|x|)}\;-\;1|\;<\;+\infty, \beta=\frac{1}{kT}.
\end{equation}

\subsection{Derivation of equations for correlation functions.}

The correlation functions of the canonical ensemble of the system $N$ of particles in some limited volume $\Lambda$ in configuration
$\eta=\{x_1,\ldots, x_m\}$ are determined by the sequence

\begin{align}\label{corfCA-2}
\rho^{(N)}_{\La}(\eta)\;:=\;\begin{cases}
 & 1,\;\;\text{at}\; \eta = \emptyset,\\
 & \varrho=\frac{N}{V}\;\;\text{at}\;\; |\eta|=1, V=\sigma(\Lambda),\\
 & 0,\;\text{at};|\eta|=m\;>\; N,
  \end{cases}
\end{align}
and for $2\leq m\leq N$,
\begin{equation}\label{corfCA-3}
\rho^{(N)}_{\La}(\eta)\;:=\;\frac{1}{Z_\Lambda(\beta, N)}\int_{\Ga^{(N-m)}_\La}\frac{d\gamma}{(N-m)!}
 e^{-\beta U(\eta\cup\ga)},
\end{equation}

where
\begin{equation}\label{SSCA}
Z_{\La}(\beta, N)\;=\;\frac{1}{N!}\int_{\Ga^{(N)}_\La}
e^{-\beta U(\ga)}d\gamma,
\end{equation}
and $\varrho$ is the density of the particles in the system.

To derive equation for correlation functions using lemma \ref{lem2}
we write down the right hand part of \eqref{corfCA-3} by the integral \eqref{IL-P}:
\begin{equation}\label{corfCA1}
\rho^{(N)}_{\La}(\eta)\;=\;\frac{1}{Z_{\La}(\beta, N)}\int_{\Ga_\La}
e^{-\beta U(\eta\cup\ga)}\1_{N-m}(\gamma)\lambda_\sigma(d\gamma), \;|\eta|\leq N,
\end{equation}
where
\begin{align}\label{unit}
\1_{N-m}(\gamma)\;:=\;\begin{cases}
 &1,\; |\gamma|=N-m,\\
 & 0,\;|\gamma|\neq N-m.
  \end{cases}
\end{align}

Introduce a notation $\eta^{(\hat{x})}:=\eta\sm\{x\}$. Let in configuration $\eta=\{x_1,..., x_m\}$ the coordinate
$x_1$ such that, for which the following inequality is true:
\begin{equation}\label{St}
  W(x_1; \eta^{(\hat{x}_1)})=\sum_{k=2}^{m}\phi(|x_1-x_k|)\;\geq\;-B,\;B\geq 0 .
\end{equation}
Then
\begin{equation}\label{SSCA}
  U(\eta\cup\ga)=W(x_1; \eta^{(\hat{x}_1)})+W(x_1; \gamma)+U(\eta^{(\hat{x}_1)}\cup\ga)
\end{equation}
and
\begin{equation}\label{corf6}
 e^{-\be W(x_1;\ga)}=\prod_{y\in\ga}\left[(e^{-\be \phi(|x_1-y|)}-1)+1\right]=\sum_{\xi\subseteq\ga}K(x_1;\xi),
\end{equation}
where
\begin{align}\label{corf7}
K(x_1;\xi)\;:=\;\begin{cases}
 &\prod_{y\in\xi}(e^{-\be \phi(|x_1-y|)}-1),|\xi|\geq 1,\\
 & 1,\;\xi=\emptyset.
  \end{cases}
\end{align}
Then
\begin{align}\label{corfCA12-1}
\rho^{(N)}_{\La}(\eta)\;=\;&\frac{e^{-\beta W(x_1; \eta^{(\hat{x}_1)})}}{Z_{\La}(\beta, N)}\int_{\Ga_\La}
e^{-\beta U(\eta^{(\hat{x}_1)}\cup\ga)} \times\notag\\
&\times\1_{N-m}(\gamma)\sum_{\xi\subseteq\ga}K(x_1;\xi)\1_{\Lambda}(\gamma\setminus\xi)\lambda_\sigma(d\gamma).
\end{align}
Apply to the integral in  \eqref{corfCA12-1} the lemma \ref{lem2} with
\begin{equation}\label{GH1}
   G(\gamma)=e^{-\beta U(\eta^{(\hat{x}_1)}\cup\ga)}\1_{N-m}(\gamma)\;\;\text{and}\;\;H(\xi;\gamma\setminus\xi)= K(x_1;\xi)\1_{\Lambda}(\gamma\setminus\xi).
\end{equation}
Then
\begin{align}\label{corfCA14}
\rho^{(N)}_{\La}(\eta)\;&=\;\frac{e^{-\beta W(x_1; \eta^{(\hat{x}_1)})}}{Z_{\La}(\beta, N)}\int_{\Ga_\La}\int_{\Ga_\La}
e^{-\beta U(\eta^{(\hat{x}_1)}\cup\xi\cup\ga)} \times\notag\\
&\times\1_{N-m}(\xi\cup\gamma)K(x_1;\xi)\1_{\Lambda}(\gamma)\lambda_\sigma(d\gamma)\lambda_\sigma(d\xi).
\end{align}
Rewrite the right hand side in the following way
\begin{align}\label{corfCA13}
&\rho^{(N)}_{\La}(\eta)=e^{-\beta W(x_1; \eta^{(\hat{x}_1)})}\frac{Z_{\La}(\beta, N-1)}{Z_{\La}(\beta, N)}\int_{\Ga_\La}K(x_1;\xi)\times\\
&\times\left[\frac{1}{Z_{\La}(\beta, N-1)}\int_{\Ga_\La}e^{-\beta U(\eta^{(\hat{x}_1)}\cup\xi\cup\ga)} \1_{N-m-|\xi|}(\gamma) \lambda_\sigma(d\gamma)\right]\lambda_\sigma(d\xi).\notag
\end{align}
Taking into account \eqref{corfCA1}($N-m-|\xi|= N-1-(m-1+|\xi|)$) we obtain:
\begin{equation}\label{corfCA123}
 \rho^{(N)}_{\La}(\eta)= e^{-\beta W(x_1; \eta^{(\hat{x}_1)})} a_N^\Lambda
 \int_{\Ga_\La}K(x_1;\xi)\rho^{(N-1)}_{\La}(\eta^{(\hat{x}_1)}\cup\xi)\lambda_\sigma(d\xi),
\end{equation}
where $|\eta|=m$\;\;$\eta=\{x_1,..., x_m\}$, $1<m<N$, and
\begin{equation}\label{coef}
 a_N^\Lambda\;=\; \frac{Z_{\La}(\beta, N-1)}{Z_{\La}(\beta, N)}.
\end{equation}

For  $m=1$
\begin{equation}\label{I-1}
 \rho^{(N)}_{\La}(\{x_1\})=\frac{N}{V}=a_N^\Lambda
 \int_{\Ga_\La}K(x_1;\xi)\rho^{(N-1)}_{\La}(\xi)\lambda_{\sigma}(d\xi),\;\;V=\sigma(\Lambda).
\end{equation}

Note, also, that
\begin{equation}\label{I-2-3}
\rho^{(N)}_\La(\eta)= 0, \;\;\text{if}\;\;  |\eta|\;\geq\;N.
\end{equation}
In the thermodynamic limit
\begin{equation}\label{lim}
\Lambda\uparrow\R^d,\;\;\lim_{\substack{V\rightarrow\infty \\
N\rightarrow\infty}}\frac{V}{N}=v,\;\;\frac{1}{v}=\varrho,
\end{equation}
 existing of which was proved in \cite{BPH69} we have
the following equation
\begin{equation}\label{limeq}
 \rho(\eta)\;=\;\varrho a(\varrho) e^{-\beta W(x_1; \eta^{(\hat{x}_1)})}
 \int_{\Ga_0}K(x_1;\xi)\rho(\eta^{(\hat{x}_1)}\cup\xi)\lambda_{\sigma}(d\xi),
\end{equation}
where
\begin{equation}\label{ratioPF}
\lim_{\substack{V=N/\varrho \\
N\rightarrow\infty}}\frac{Z_\Lambda(\beta, N-1)}{Z_\Lambda(\beta, N)}\;=\;\varrho a(\varrho),\;\;V=\sigma(\Lambda)
\end{equation}
and
\begin{equation}\label{ratioPF1}
 a(\varrho)=\frac{1}{\w{Q}(\rho)}, \;\w{Q}(\rho)\;=\;\int_{\Ga_0}K(x_1;\xi)\rho(\xi)\lambda_{\sigma}(d\xi).
\end{equation}

\section{ Construction of expansion for correlation functions in the density parameter $ \varrho $.}

Solution of the equation \eqref {limeq} by the iteration method (Ruelle operator method \cite {Ru69}) or
by the Minlos-Poghosyan method \cite {MP77} (about the connection between these methods see \cite {Re2021})
leads to the series in the parameter $ \varrho a(\varrho) $. So, in fact, these equations are nonlinear,
because $ a(\varrho) $ depends on all correlation functions (see \eqref{limeq} and \eqref{ratioPF1}).
In this section, we construct expansion in the  parameter density $ \varrho $, considering the equation \eqref{limeq}
as nonlinear equation.

\subsection{Solution of equations for correlation functions.}

Rewrite the equation \eqref{limeq}, \eqref{ratioPF1} as follows:

\begin{equation}\label{limeq-1}
 \w{Q}(\rho_j)\rho_j(\eta)\;=\;\varrho  e^{-\beta W(x_1; \eta^{(\hat{x}_1)})}
 j(x_1)\int_{\Ga_0}K(x_1;\xi)\rho_j(\eta^{(\hat{x}_1)}\cup\xi)\lambda_{\sigma}(d\xi),
\end{equation}
\begin{equation}\label{ratioPF1-1}
 \w{Q}(\rho_j)\;=\;\int_{\Ga_0}K(x_1;\xi)\rho_j(\xi)\lambda_{\sigma}(d\xi),
\end{equation}
where, following \cite{MP77}, a continuous  positive function
$ j: \R^d \rightarrow \R_+ $ is introduced and $ \rho (\eta) $ becomes functional
such as $\rho(\eta)\;=\;\rho_j(\eta)|_{j=1}.$

We will look for the solution of these equations in the form:
\begin{equation}\label{SSCA-2}
\rho_j(\eta)\;=\;e(j;\eta)\int_{\Gamma^0}T(\eta|\gamma)e(j;\ga)\lambda_\sigma(d\gamma),
\end{equation}
where $e(j;\eta)$ is given by \eqref{I-lam4}.

The idea of finding a solution is to consider the left side of the equation \eqref {limeq-1} as the product of two generating
functionals (see \eqref{odprod}) and compare it with the right part, which is also written in the form of a generating functional
of the form \eqref {GFf}. So, substituting in \eqref {ratioPF1-1} the corresponding expression for $ \rho_j (\xi) $ we have:
\begin{equation}\label{ratioPF1-1a}
 \w{Q}(\rho_j)\;=\;\int_{\Ga_0}\int_{\Ga_0}e(j;\xi\cup\ga)K(x_1;\xi)T(\xi|\gamma)\lambda_{\sigma}(d\gamma)\lambda_{\sigma}(d\xi).
\end{equation}
Apply to the integral on the right side of the \eqref {ratioPF1-1a} lemma \ref {lem2} with
\begin{equation}\label{GH1}
   G(\xi\cup\gamma)=e(j;\xi\cup\ga)\;\;\text{and}\;\;H(\xi;\gamma)= K(x_1;\xi)T(\xi|\gamma).
\end{equation}
Then
\begin{equation}\label{ratioPF1-2}
 \w{Q}(\rho_j)\;=\;\w{F}_{\psi_1}(j)\;=\;\int_{\Ga_0} e(j;\ga)\psi_1(\gamma)\lambda_{\sigma}(d\gamma),
\end{equation}
with
\begin{equation}\label{ratioPF1-2a}
 \psi_1(\gamma)\;=\;\sum_{\xi\subseteq\gamma}K(x_1;\xi)T(\xi|\gamma\setminus\xi).
\end{equation}
Write down  \eqref{SSCA-2} in the form of generating functional:
\begin{equation}\label{SSCA-3}
\rho_j(\eta)\;=\; \w{F}_{\psi_2}(j)\;=\;\int_{\Ga_0} e(j;\ga)\psi_2(\gamma)\lambda_{\sigma}(d\gamma),
\end{equation}
with
\begin{equation}\label{ratioPF1-2b}
 \psi_2(\gamma)\;=\;e(j;\eta) T(\eta|\gamma).
\end{equation}
Left hand side \eqref{limeq-1} is:
\begin{equation}\label{limeq-1-left}
 \w{Q}(\rho_j)\rho_j(\eta)\;=\; \w{F}_{\psi_1}(j)\w{F}_{\psi_2}(j)\;=\;\w{F}_{\psi_1*\psi_2}(j)\;
 =\;\int_{\Ga_0} e(j;\ga)(\psi_1*\psi_2)(\gamma)\lambda_{\sigma}(d\gamma),
\end{equation}
where
\begin{align}\label{help}
(\psi_1*\psi_2)(\gamma)&=\sum_{\xi\subseteq\gamma}\psi_1(\xi)\psi_2(\gamma\setminus\xi)=\sum_{\xi\subseteq\gamma}\psi_2(\gamma\setminus\xi)\psi_1(\xi)=\\
&=e(j;\eta)\sum_{\xi\subseteq\gamma} T(\eta|\gamma\setminus\xi)\sum_{\zeta\subseteq\xi}K(x_1;\zeta)T(\zeta|\xi\setminus\zeta)\notag
\end{align}

Right hand side \eqref{limeq-1} after substitution of $\rho_j(\eta^{(\hat{x}_1)}\cup\xi)$
in the form \eqref{SSCA-2} will look as:
\begin{equation}\label{limeq-1-right0}
\varrho  e^{-\beta W(x_1; \eta^{(\hat{x}_1)})}
 e(j;\eta)\int_{\Ga_0}\int_{\Ga_0}e(j;\xi\cup\ga)K(x_1;\xi)T(\eta^{(\hat{x}_1)}\cup\xi|\gamma)\lambda_{\sigma}(d\gamma)
\lambda_{\sigma}(d\xi).
\end{equation}
Let us use lemma \ref{lem2} again with
\begin{equation}\label{GH1}
   G(\xi\cup\gamma)=e(j;\xi\cup\ga)\;\;\text{and}\;\;H(\xi;\gamma)= K(x_1;\xi)T(\eta^{(\hat{x}_1)}\cup\xi|\gamma).
\end{equation}
Then right hand side of \eqref{limeq-1} will be as
\begin{equation}\label{limeq-1-right}
\varrho  e^{-\beta W(x_1; \eta^{(\hat{x}_1)})}
 e(j;\eta)\int_{\Ga_0} e(j;\ga)\sum_{\xi\subseteq\gamma} K(x_1;\xi)T(\eta^{(\hat{x}_1)}\cup\xi|\gamma\setminus\xi)\lambda_{\sigma}(d\gamma).
\end{equation}
We equate the functions near $ e(j;\ga) $ under the integrals of the expressions \eqref {limeq-1-left} and \eqref {limeq-1-right} and, reducing the multiplier
   $ e(j; \eta) $ we get the relation:
\begin{align}\label{help}
 &\sum_{\xi\subseteq\gamma} T(\eta|\gamma\setminus\xi)\sum_{\zeta\subseteq\xi}K(x_1;\zeta)T(\zeta|\xi\setminus\zeta)\;=\\
 &=\;\varrho  e^{-\beta W(x_1; \eta^{(\hat{x}_1)})}\sum_{\xi\subseteq\gamma} K(x_1;\xi)T(\eta^{(\hat{x}_1)}\cup\xi|\gamma\setminus\xi).\notag
\end{align}
Select the term with $ \xi = \emptyset $ in the left side of this relation and take into account that $ T(\emptyset | \emptyset) = 1 $ and $ T (\emptyset | \gamma) = 0 $,
if $ \gamma \neq \emptyset) $, we obtain the following nonlinear recurrent relation:
\begin{align}\label{final}
T(\eta|\gamma)&= \varrho  e^{-\beta W(x_1; \eta^{(\hat{x}_1)})}\sum_{\xi\subseteq\gamma} K(x_1;\xi)T(\eta^{(\hat{x}_1)}\cup\xi|\gamma\setminus\xi)-\notag\\
&-\sum_{\emptyset\neq\xi\subseteq\gamma} T(\eta|\gamma\setminus\xi)\sum_{\emptyset\neq\zeta\subseteq\xi}K(x_1;\zeta)T(\zeta|\xi\setminus\zeta).
\end{align}

{\bf  Initial conditions:}
\begin{equation}\label{incon0}
    T(\emptyset|\emptyset)\;=\;1,\;\;\;T(\emptyset|\gamma)\;=\;0, \text{if}\;\;\gamma\neq\emptyset,
\end{equation}
\begin{equation}\label{incon1}
    T(\{\eta|\gamma)\;=\;0,\text{if}\;\;\eta\cap\gamma\neq\;\emptyset.
\end{equation}

\subsection{Graphic interpretation of solutions.}

 As in the case of the Kirkwood-Salzburg equations for a large canonical ensemble (see
   details in \cite {Re2021}) the analytical expression of the kernel $ T (\eta|\gamma) $ can be given as
  contributions of forest-graphs.   The number $ N (m | n), m = |\eta|, n = |\gamma| $
all forest graphs can be calculated by solving the following recurrent equation:
\begin{equation}\label{Numberforests-vir}
N(m|n)=\sum_{k=0}^n {n \choose k}N(m+k-1|n-k)+\sum_{k=1}^n {n \choose k}N(m|n-k)\sum_{l=1}^k {k \choose l}N(l|k-l).
\end{equation}
If we reject the nonlinear term, the recurrent equation can be solved exactly: $ N (m | n) = m (m + n)^{n-1} $ (see \cite {Re2021}).
This formula makes it possible to prove the convergence of the integral \eqref {SSCA-2} at $ j = 1 $ and small density values.
But the nonlinear term apparently significantly increases the number of forest graphs. But according to the formula \eqref {final},
each contribution of the forest graph, which contains a tree with the contribution of
the factor $ K (x_1; \xi), \xi \in \gamma $, which appears from the second sum in \eqref {final} is included with a minus sign. So, for example calculating
\begin{equation}\label{T11}
    T(x_1|y)\;=\; \varrho^2\left[ K(x_1;y)\;-\;K(x_1;y)\right]=0.
\end{equation}
Therefore, it is necessary to analyze this recurrent relationship in more detail.
It is easy to get the following proposition.

\begin{prop}\label{dens}
From initial conditions \eqref{incon0}-\eqref{incon1} it follows that
\begin{equation}\label{den1}
   T(x_1|\emptyset)\;=\;\varrho,\;\;T(x_1|\gamma)\;=\;0\;\text{for}\;\gamma\neq\emptyset.
\end{equation}
It gives the needed equality for density $\rho(x_1)\;=\;\varrho$. \\
In addition   $T(\eta|\emptyset)\;=\;\varrho^{|\eta|}\exp[-\beta U(\eta)]$.
\end{prop}
The {\it proof} follows from \eqref{final} by induction procedure.
To obtain the solution of \eqref{final} in the language of graph theory \cite {Ste64, Gro67}
define the set graphs $\mathcal{D}(\eta; \gamma)$.

\begin{remark}\label{graph}
The set  $\mathcal{D}(\eta; \gamma)$ is the set of rooted graph,
connected components of which are 2-connected graphs with respect to configuration $\gamma$.
Any graph $G\in\mathcal{D}(\eta; \gamma) $ has $m=|\eta|$ vertices of configuration $\eta$ and $n=|\gamma|$ vertices of
configuration $\gamma$. Every vertex of $\eta$ can be free of lines or can connect only with vertices $y$ of configuration
$\gamma$. Every vertex $y\in\gamma$ connects with vertices $x$ or $y$ by at least two lines.
This definitions is very close to definitions of the set of white and black vertices $\mathcal{D}(W; B)$ in \cite {Ja21}, Section 2.
The role of the roots of the graph is performed by the configuration $\eta$.
A singleton $G\in\mathcal{D}(x_1; \emptyset)$.  The contribution of any vertex is $\varrho$. The contribution of line $w(l_{xy})$,
which connected two vertices $x$ and $y$ or $y_i$ and $y_j$ is $K(x;y)$.
\end{remark}

\begin{remark}\label{graph1}
According to \eqref{SSCA-2}, integration is performed at the points of configuration $\gamma$, and according to equation \eqref{final},
 in the kernels $T(\theta; \xi)$, the configuration $\xi\subseteq\gamma$, and the configuration $\theta$ can contain both variables
 from $\eta$ and $\gamma$. Therefore, the corresponding graphs of sets $\mathcal{D}(\theta; \xi)$ can contain only black vertices.
\end{remark}

The main statement  of the section is

\begin{prop}\label{dens}
The solution of  the recurrence relation \eqref{final}  can be rewritten by contributions from the graphs of the set $\mathcal{D}(\eta; \gamma)$:
\begin{equation}\label{den1}
   T(\eta|\gamma)\;=\;\varrho^{|\eta|+|\gamma|}\exp[-\beta U(\eta)] \sum_{G\in\mathcal{D}(\eta; \gamma) }w_G(L; \eta, \gamma),
\end{equation}
where $L$ is a configuration of lines(edges).
\end{prop}
The main points of {\it proof}. The proof is based on the induction procedure.
It is very easy to see the cancellation of the contributions of the corresponding graphs, for example, for kernels
$T(x_1, x_2|y_1)$ and $T(x_1, x_2|y_1, y_2)$. To facilitate these calculations, it is better to consider instead of the equation
\eqref{final} the equation for kernels that are integrated in variables $y_1,..., y_n$:

\begin{equation}\label{kern2}
   \widehat{T}(\eta|n):=\int_{\Gamma^{(n)}} T(\eta|\gamma)d\gamma =\int_{\R^{dn}} (dy)^nT(\eta|\{y\}_1^n),
\end{equation}
especially since this is the value that must be substituted into the formula for the solution \eqref{SSCA-2}.
\begin{align}\label{final2}
\widehat{T}(\eta|n)&=\varrho  e^{-\beta W(x_1; \eta^{(\hat{x}_1)})}\sum_{k=0}^n {n \choose k}
\int (dy)^k K(x_1;\{y\}_1^k)\widehat{T}(\eta^{(\hat{x}_1)}\cup\{y\}_1^k)|n-k)-\notag \\
&-\sum_{k=1}^n {n \choose k}\sum_{l=1}^k {k \choose l}\int (dy)^l K(x_1;\{y\}_1^l)\widehat{T}(\eta|n-k)\widehat{T}(\{y\}_1^l|k-l).
\end{align}
For $k=0$ in the first line of formula \eqref{final2}, the term $\widehat{T}(\eta^{(\hat{x}_1)}|n)$ corresponds to the analytical
contribution of all graphs $\mathcal{D}(\eta; \gamma)$, in which the vertex $x_1$ has no edges.
For $k\geq 1$ the set of graphs for $\widehat{T}(\eta^{(\hat{x}_1)}\cup\{y\}_1^k)|n-k)$ is arranged in such a way
that $k$ vertices from $n$ vertices of the configuration $\gamma$ are transformed into vertices, which are connected
to the graphs of $\widehat{T}(\eta^{(\hat{x}_1)}|n-k)$ either by means of lines or as free vertices.
If a vertex $y_i$ is already connected by at least one line, then joining a line with vertex $x_1$
leaves this graph in the set $\mathcal{D}(\eta; \gamma)$. If the vertex did not have edges,
then operation  $\int (dy)^l K(x_1;\{y\}_1^l)$ does not ensure that it belongs to the set
$\mathcal{D}(\eta; \gamma)$, and such a diagram should be reduced.This is ensured by the presence of
a similar contribution in the second line of the formula \eqref{final2}.
This variant occurs in $l$ of $k$ cases for  each of the graphs in  $\widehat{T}(\eta|n-k)$.
\hfill $\blacksquare$

\subsection{ Conclusion on the convergence of the expansion.}

Existence of a unique solution of the nonlinear equation \eqref{limeq} as an operator equation in the Banach space $ E_\xi $
was proved in \cite{Po75}. In a recent work of Jansen \cite{Ja21} a generalized version of Groeneveld’s convergence criterion for the
virial expansion and generating functionals for weighted 2-connected graphs was proven. In fact, the present paper turned out 
to be very close to article \cite{Ja21}, and here one can apply the criteria for the convergence of expansions.
Therefore, we do not yet discuss this issue in this preliminary publication. Although perhaps a more thorough study 
of the recurrence relation \eqref{final2} may introduce some adjustments.

\vspace{0.2cm}

{\bf AVAILABILITY OF DATA}

\vspace{0.2cm}

Data available on request from the
author

\vspace{0.2cm}

\small{}

\end{document}